\documentclass[sigconf]{acmart}
\usepackage{multirow}
\AtBeginDocument{%
  }

\setcopyright{acmlicensed}
\copyrightyear{2018}
\acmYear{2018}
\acmDOI{XXXXXXX.XXXXXXX}
\acmConference[Conference acronym 'XX]{Make sure to enter the correct
  conference title from your rights confirmation email}{June 03--05,
  2018}{Woodstock, NY}
\acmISBN{978-1-4503-XXXX-X/2018/06}

\begin{document}

\title{COINS: SeManti\textbf{\textcolor{red}{C}} Ids Enhanced C\textcolor{red}{\textbf{O}}ld \textcolor{red}{\textbf{I}}tem Representatio\textcolor{red}{\textbf{N}} for Click-through Rate Prediction in E-commerce \textcolor{red}{\textbf{S}}earch} 

\author{Qihang Zhao}
\email{zhaoqh75@mail.ustc.edu.cn}
\affiliation{%
  \institution{Kuaishou Inc.}
  \city{Hangzhou}
  \state{Zhejiang}
  \country{China}
}

% \author{Qihang Zhao, Zhongbo Sun, Xiaoyang Zheng, Xian Guo, Siyuan Wang, Zihan Liang, Mingcan Peng, Ben Chen*, Chenyi Lei}
% \authornote{*corresponding author}
% \email{zhaoqh75@mail.ustc.edu.cn, sunzb17@gmail.com, {zhengxiaoyang, guoxian, wangsiyuan, liangzihan, pengmingcan, chenben}@kuaishou.com, leichy@mail.ustc.edu.cn}
% \affiliation{%
%   \institution{Kuaishou Inc.}
%   \city{Hangzhou}
%   \state{Zhejiang}
%   \country{China}
% }

\author{Zhongbo Sun}
\affiliation{%
  \institution{Kuaishou Technology}
  \city{Hangzhou}
  \country{China}}
\email{sunzb17@gmail.com}

\author{Xiaoyang Zheng}
\affiliation{%
  \institution{Kuaishou Technology}
  \city{Hangzhou}
  \country{China}}
\email{zhengxiaoyang@kuaishou.com}

\author{Xian Guo}
\affiliation{%
  \institution{Kuaishou Technology}
  \city{Beijing}
  \country{China}}
\email{guoxian@kuaishou.com}

\author{Siyuan Wang}
\affiliation{%
  \institution{Kuaishou Technology}
  \city{Beijing}
  \country{China}}
\email{wangsiyuan@kuaishou.com}

\author{Zihan Liang}
\affiliation{%
  \institution{Kuaishou Technology}
  \city{Hangzhou}
  \country{China}}
\email{liangzihan@kuaishou.com}

\author{Mingcan Peng}
\affiliation{%
  \institution{Kuaishou Technology}
  \city{Beijing}
  \country{China}}
\email{pengmingcan@kuaishou.com}

\author{Ben Chen}
\authornote{*corresponding author}
\affiliation{%
  \institution{Kuaishou Technology}
  \city{Hangzhou}
  \country{China}}
\email{chenben03@kuaishou.com}

\author{Chenyi Lei}
\affiliation{%
  \institution{Kuaishou Technology}
  \city{Hangzhou}
  \country{China}}
\email{leichy@mail.ustc.edu.cn}

\renewcommand{\shortauthors}{Zhao et al.}

\begin{abstract}
  With the rise of modern search and recommendation platforms, insufficient collaborative information of cold-start items exacerbates the Matthew effect of existing platform items, challenging platform diversity and becoming a longstanding issue. Existing methods align items' side content with collaborative information to transfer collaborative signals from high-popularity items to cold-start items. However, these methods fail to account for the asymmetry between collaboration and content, nor the fine-grained differences among items. To address these issues, we propose COINS, an item representation enhancement approach based on fused alignment of semantic IDs. Specifically, we use RQ-OPQ encoding to quantize item content and collaborative information, followed by a two-step alignment: RQ encoding transfers shared collaborative signals across items, while OPQ encoding learns items' differentiated information. Comprehensive offline experiments on large-scale industrial datasets demonstrate COINS’s superiority, and rigorous online A/B tests confirm statistically significant improvements: item CTR +1.66\%, buyers +1.57\%, and order volume +2.17\%.
\end{abstract}

\begin{CCSXML}
<ccs2012>
 <concept>
  <concept_id>00000000.0000000.0000000</concept_id>
  <concept_desc>Do Not Use This Code, Generate the Correct Terms for Your Paper</concept_desc>
  <concept_significance>500</concept_significance>
 </concept>
 <concept>
  <concept_id>00000000.00000000.00000000</concept_id>
  <concept_desc>Do Not Use This Code, Generate the Correct Terms for Your Paper</concept_desc>
  <concept_significance>300</concept_significance>
 </concept>
 <concept>
  <concept_id>00000000.00000000.00000000</concept_id>
  <concept_desc>Do Not Use This Code, Generate the Correct Terms for Your Paper</concept_desc>
  <concept_significance>100</concept_significance>
 </concept>
 <concept>
  <concept_id>00000000.00000000.00000000</concept_id>
  <concept_desc>Do Not Use This Code, Generate the Correct Terms for Your Paper</concept_desc>
  <concept_significance>100</concept_significance>
 </concept>
</ccs2012>
\end{CCSXML}

\ccsdesc[500]{Information systems~Content ranking}

\keywords{Semantic Ids; Cold Start}

\received{29 September 2025}

\maketitle

\section{Introduction}
E-commerce search recommendation systems have become core channels for connecting users and commodities, with continuous optimization in representation learning and matching efficiency to adapt to the rapid growth of platform scale. However, the item cold-start challenge has become increasingly prominent—newly added commodities (accounting for approximately 30\% of monthly updates on our platforms) often face extreme sparsity of user interaction data. Over 60\% of these cold-start items receive fewer than 5 clicks in their first week, lacking critical collaborative signals (e.g., user click preferences, purchase tendencies) that traditional recommendation methods rely on. This directly degrades search recommendation accuracy for cold items, making it an urgent need to develop efficient cold-start solutions compatible with existing industrial systems.

Existing cold-start solutions fall into two paradigms. Generator-based methods \cite{10.1145/3477495.3531897,10.1145/3690624.3709336,10.1007/978-981-97-2262-4_9,10.5555/3295222.3295434} synthesize cold-item representations via warm-item signals: meta-learning \cite{10.1145/3690624.3709336} pre-trains transferable generators for few-sample adaptation, GANs \cite{10.1145/3477495.3531897} align cold embeddings with warm signal distributions, and VAEs \cite{10.1007/978-981-97-2262-4_9} sample from learned warm-data distributions. Knowledge alignment-based methods \cite{10.1145/3696410.3714852,10.1145/3626772.3657839,10.1145/3711858,10.1145/3474085.3475665} bridge content and collaborative signals: contrastive learning optimizes \cite{10.1145/3474085.3475665} content encoders to match warm collaborative representations, while knowledge distillation \cite{10.1145/3696410.3714852} uses content features as a medium to transfer teacher-model (warm-item) knowledge to student models (cold-item).

\begin{figure}
  \includegraphics[width=0.5\textwidth]{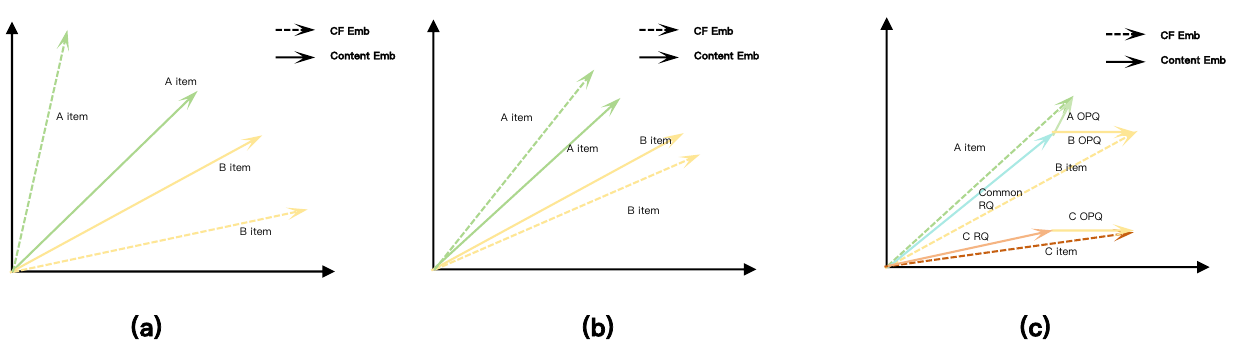}
  \caption{Types of  Knowledge alignment-based methods.}
  \Description{Enjoying the baseball game from the third-base
  seats. Ichiro Suzuki preparing to bat.}
  \label{fig:aligh_demo}
\end{figure}

Despite the partial success of existing approaches, two critical limitations persist. First, for generative models: few-shot-based generators independently generate representations for cold-start items, yet they overlook temporal dynamics and temporal variations during feature transfer—this flaw readily induces distribution drift in the generated cold-start item embeddings. Second, for knowledge alignment-based methods: such approaches generally rely on the core assumption that item’s side content information and collaborative signals are consistent in the embedding space. Even though some improved variants attempt to learn the consistency between these two components prior to alignment, they still do not deviate from the knowledge alignment framework. Their core limitation remains in neglecting the unique fine-grained discriminative information of individual items, ultimately confining recommendation performance to a suboptimal level while failing to address the asymmetry between side information and collaborative signals that frequently arises in real-world scenarios, as illustrate in Figure \ref{fig:aligh_demo} (a) and (b).

To alleviate these issues, this paper proposes a novel item representation fusion and enhancement method based on semantic IDs, termed COINS. Specifically, inspired by OneSearch \cite{chen2025onesearchpreliminaryexplorationunified}: a recently deployed end-to-end generative framework that has achieved successful application in e-commerce search. We leverage its proposed RQ-OPQ encoding (endowed with rich hierarchical semantic and collaborative signals) to enhance cold-start item representations. First, for the coarser-grained RQ encoding, we design an adaptive transfer mechanism to enable bidirectional transfer between high-frequency collaborative signals embedded in item IDs and dense semantic information within RQ encodings. Second, for the residual vectors of items obtained via RQ hierarchical quantization (i.e., OPQ encodings), which encapsulate the discriminative information of different items, we integrate such discriminative information into item representations through contrastive learning. Ultimately, the resulting item representations not only incorporate abundant content and collaborative information but also accurately capture the uniqueness of individual items, as illustrate in Figure \ref{fig:aligh_demo} (c). The main contributions are as follows:
\begin{itemize}
    \item We propose an innovative adaptive transfer mechanism to achieve adaptive bidirectional alignment between the collaborative information embedded in item IDs and the semantic information within RQ encodings;
    \item We are the first to propose an OPQ encoding-based learning strategy for item discriminative information, enabling cold-start item representations to accurately capture their unique attributes;
    \item We conduct comprehensive offline experiments and rigorous online A/B test, whose results fully validate the superiority of our proposed COINS method.
\end{itemize}

\section{Methodology}
\subsection{Cold Start for Search}
Typically, the core of the cold-start problem lies in exploring how to predict user interaction behaviors for new items. In this work, we focus on the click-through rate (CTR) prediction task and investigate how to improve the performance of cold-start tasks through the synergy between Semantic IDs generated by large search models and collaborative signals. In this section, we first formally define the cold-start problem, then elaborate on the proposed COINS framework in detail. Formally, given a user U (user profile $U_p$), the user’s current search query Q, and the user’s historical behavior sequence \(H_U\), cross features $C$, each sample in the CTR prediction task can be represented as \(I = (id, X)\), which includes a label \(y \in \{0,1\}\) (indicating whether the sample is clicked by the user): id denotes the random hash ID of the item, and $X$ represents the multi-dimensional feature set of the item (including content features, metadata features, etc.). The goal of cold-start CTR estimation is to learn a discriminative model $f$ to output the click probability of the sample:
\begin{equation}
    \label{equ_f}
    \hat{y} = f(id, X, Q, H_U, U_p, C; \theta)\
\end{equation}
where \(\theta\) denotes the parameters of model f. The above model is optimized via the binary cross-entropy (BCE) loss:
\begin{equation}
    \label{equ_loss_bce}
    \mathcal{L}_{BCE} = -\frac{1}{N}\sum_{i=1}^N \left[ y_i \log \hat{y}_i + (1-y_i) \log (1-\hat{y}_i) \right]\
\end{equation}
Specifically, compared with warm items (items with sufficient interaction data), cold-start items suffer from scarce user interaction records, leading to the absence of statistical features (e.g., click frequency, conversion rate) and partial ID-related representations. This further results in difficulties in model convergence during training.
\subsection{RQ-OPQ Encodings}
RQ-OPQ encoding is a core technical innovation proposed in Kuaishou’s end-to-end generative framework OneSearch for e-commerce search. This encoding method combines the advantages of Residual Quantization (RQ) and Optimized Product Quantization (OPQ), modeling item features from two dimensions: vertically (hierarchical semantics) and horizontally (unique features). Specifically, its implementation process is as follows: First, a two-tower vector retrieval model optimized for content features and conversion signals is trained to generate the initial embedding of each item. Second, based on this initial embedding, the RQ-Kmeans algorithm is applied to perform three-layer hierarchical encoding, which primarily focuses on capturing the core attributes and shared semantic information of items. Third, for the residual information remaining after RQ encoding, the OPQ algorithm is used to conduct two-layer quantization encoding—this step is specifically designed to capture and quantize the differentiated fine-grained features of items. Finally, each item is represented by a five-layer semantic ID (sid): $I_{sid}=(RQ_1, RQ_2, RQ_3, OPQ_1, OPQ_2)$.
\subsection{Semantic Ids Enhanced Item Representation}
\begin{figure}
  \includegraphics[width=0.5\textwidth]{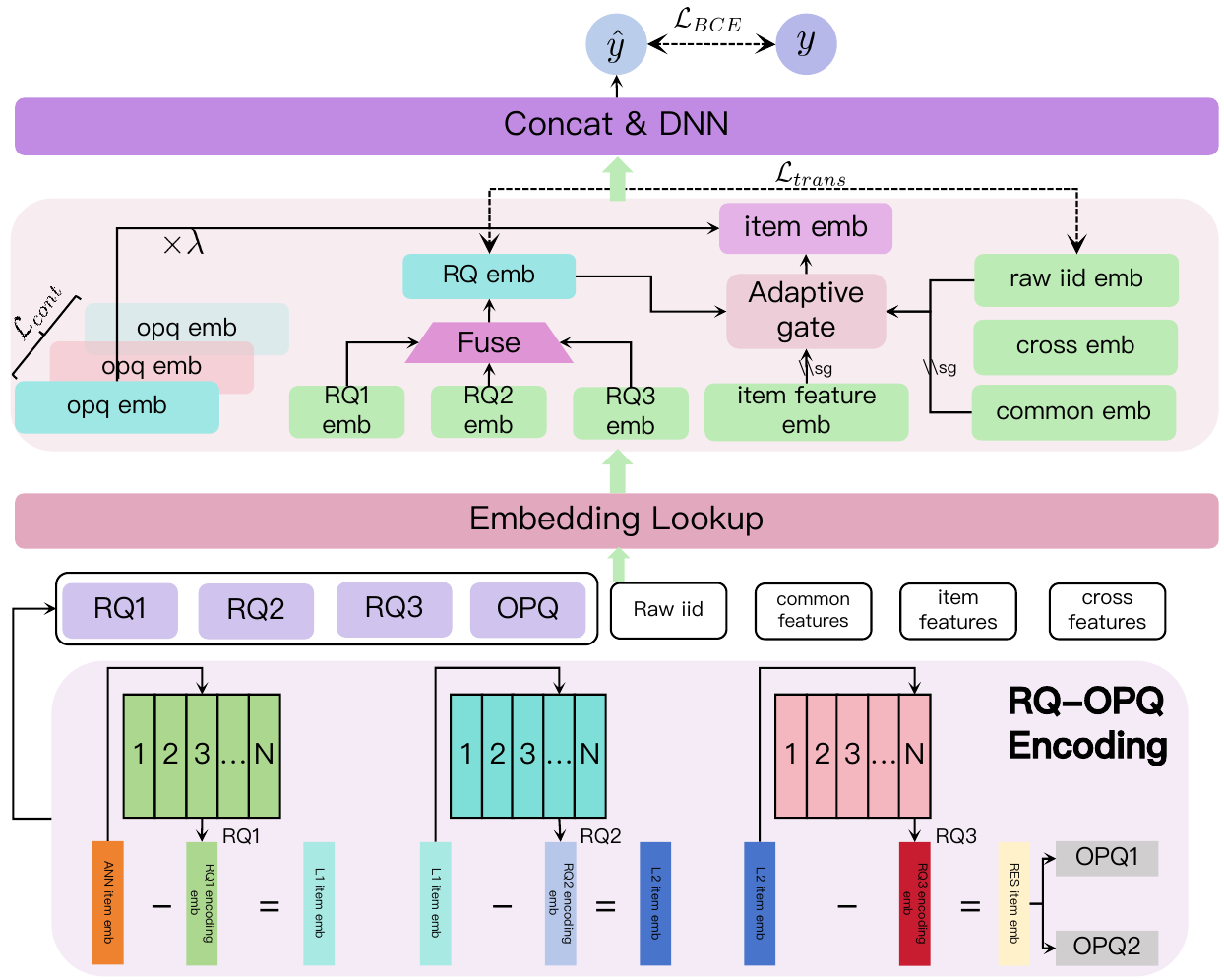}
  \caption{The framework of COINS.}
  \label{fig:smile_framework}
\end{figure}

In this section, we elaborate in detail on COINS, a cold-start item representation enhancement framework based on semantic IDs. This framework comprises two core modules: the first is an adaptive transfer and alignment mechanism between RQ encoding and item IDs, which aims to learn coarse-grained representations with both semantic information and collaborative signals; the second is an item discriminative information learning module based on OPQ encoding, through which fine-grained discriminative information is injected into item representations, as illustrated in Figure \ref{fig:smile_framework}.

\subsubsection{Adaptive Information Transfer via RQ Encoding}
RQ encoding carries convergent semantic information of items, and different items may share partial codebooks. Meanwhile, the random hash IDs of items have accumulated abundant collaborative signals during the continuous incremental training of the original CTR model. Therefore, we perform bidirectional transfer between the collaborative signals carried by item IDs and the semantic information contained in RQ encoding. Specifically, we first map the RQ encoding of items and item IDs to the same dimensional space, laying the foundation for subsequent information transfer:
\begin{equation}
    \label{equ_fuse_rq}
    I_{emb}^{RQ} = Fuse(RQ_1,RQ_2,RQ_3)
\end{equation}
Notably, warm items have sufficient user interaction records, while the interaction data of cold-start items is extremely sparse. Thus, when performing item representation fusion and alignment, we design an adaptive transfer mechanism: specifically, we input context features, user profiles, and partial conversion features of items into an adaptive transfer gate network, which outputs corresponding weights $\mathcal{T}_g$ to achieve adaptive fusion and alignment between item IDs and item RQ codebooks. Additionally, we dynamically determine the effective weights of different representations through context $C$, user profile $U_p$, and item features $X$, ultimately generating an item representation suitable for the current request:
\begin{align}
    \label{equ_adaptive_gate}
    \mathcal{T}_g &= DNN(C, U_p, X) \\
    I_{emb}^{c} &= \mathcal{T}_g *I_{emb}^{id} + (1-\mathcal{T}_g)*I_{emb}^{RQ}
\end{align}
Meanwhile, the entire transfer process is constrained by the following loss function: this process also introduces a transfer gate to ensure the directionality of information transfer. Specifically, for warm item representations, this mechanism tends to preserve the collaborative signal distribution of item IDs while transferring their collaborative information to RQ representations; for cold-start item representations, it injects the collaborative signals obtained through convergent transfer in RQ into item IDs, while preserving the semantic information distribution of their RQ encoding:
\begin{equation}
    \label{equ_loss_trans}
    \mathcal{L}_{trans} = \mathcal{T}_g * KL(sg(I_{emb}^{id}), I_{emb}^{RQ}) + (1-\mathcal{T}_g)KL(I_{emb}^{id} ,sg(I_{emb}^{RQ}))
\end{equation}
where $KL$ denotes KL divergence, $sg$ denotes stop gradient operation.

\subsubsection{Enhancing Item Differentiated Information with OPQ Encoding}
OPQ encoding captures the differentiated features of items after hierarchical quantization, and can inject fine-grained information into item representations to highlight the uniqueness of different items. To further enhance the differentiated information representation capability of OPQ encoding, we introduce a contrastive learning strategy. 
First, based on the similarity of OPQ codebook vectors, we select the Top-10 similar OPQ encodings as candidate positive OPQ encodings for each OPQ encoding. During the training phase, for any item $i$, items with candidate positive OPQ encodings of item $i^+$ are filtered from the current training batch and designated as positive samples; meanwhile, the remaining items and additional randomly selected items $i^\_$ are used as negative samples to ensure significant differences between positive and negative samples. We optimize OPQ encoding using the contrastive InfoNCE loss function. Assuming the positive sample set of item $i$ is $P$, the negative sample set is $N$, the contrastive loss function is defined as:
% \begin{equation}
%     \label{equ_loss_cont}
% \mathcal{L}_{cont} = -\log\left( \frac{\sum_{i^+ \in P} \exp\left( \text{sim}(I_{emb}^{i_{OPQ}}, I_{emb}^{i^+_{OPQ}}) / \tau \right)}{\sum_{i^+ \in P} \exp\left( \text{sim}(I_{emb}^{i_{OPQ}}, I_{emb}^{i^+_{OPQ}}) / \tau \right) + \sum_{i^- \in N} \exp\left( \text{sim}(I_{emb}^{i_{OPQ}}, I_{emb}^{i^-_{OPQ}}) / \tau \right)} \right)
% \end{equation}

\begin{equation}
    \label{equ_loss_cont}
\mathcal{L}_{cont} = -\log\left( \frac{\sum_{i^+ \in P} \exp\left( \text{sim}(I^{i}, I^{i^+}) / \tau \right)}{\sum_{i^+ \in P} \exp\left( \text{sim}(I^{i}, I^{i^+}) / \tau \right) + \sum_{i^- \in N} \exp\left( \text{sim}(I^{i}, I^{i^-}) / \tau \right)} \right)
\end{equation}

where $sim(.)$ is OPQ embedding consine similarity of item pair , $\tau$ is the temperature parameter.
Finally, we obtained the item representation $I_{emb}^{f}$ to replace the $I^{id}_{emb}$ in the subsequent pipeline:
\begin{equation}
    \label{equ_final_item}
I_{emb}^f = I_{emb}^c + \lambda * I_{emb}^{OPQ}
\end{equation}

The entire model is optimized through the following loss function:
\begin{equation}
    \label{equ_loss_total}
\mathcal{L}_{total} = \mathcal{L}_{BCE} + \alpha _1*\mathcal{L}_{trans} + \alpha _2*\mathcal{L}_{cont}
\end{equation}

\section{Experiment}
\subsection{Dataset}
We collected 91 days of real user logs from an online e-commerce platform, with the first 90 days for training and the last day for testing. The test set contains 500 million samples, and items are categorized by the following criteria: warm items are defined as those with clicks >3 or orders >0 within 7 days; cold-start items are those with impressions <200 within 7 days. This results in 400 million cold-start samples and 100 million warm samples, whose distribution aligns with the Pareto Principle — 20\% of items account for 80\% of the traffic.
\subsection{Experiment Setup}
\subsubsection{Metrics} We use AUC and GAUC to evaluate the performance of CTR ranking models.
\subsubsection{Implementation Details} To ensure experimental reproducibility, the hyperparameter settings mentioned in the method section are listed below: Hyperparameters$\alpha_1=0.01$ and $\alpha_2=0.05$ are set to control the consistency of the numerical magnitudes of the three loss functions; the temperature coefficient $\tau=0.1$ is used to ensure distribution smoothness; the OPQ representation fusion coefficient $\lambda$ is set to 0.5.
\subsubsection{Baselines}Since our proposed COINS framework is based on semantic ids to enhance the representation of cold start items, we have selected some models that are also based on semantic ids or cold start as our baseline:

\textbf{SPM\_SID \cite{DBLP:conf/recsys/SinghV0KSZHHWTC24}:} hashing sub fragments of semantic ids sequences to adaptively obtain the semantics of different items.

\textbf{DAS \cite{DBLP:journals/corr/abs-2508-10584}:} Through joint training of semantic collaborative models, use contrastive learning to synchronously optimize the quantization and alignment process of semantic ids and item id.

\textbf{SaviorRec \cite{yao2025saviorrec}:} A cold start model based on multimodal information and semantic ids collaborative training to enhance item representation.
\subsection{Offline Performance}

\begin{table}[]
\caption{The offline performance of COINS.}
\begin{tabular}{ccccccc}
\hline
          & \multicolumn{2}{c}{All}           & \multicolumn{2}{c}{Warm}          & \multicolumn{2}{c}{Cold}          \\ \cline{2-7} 
          & AUC             & GAUC            & AUC             & GAUC            & AUC             & GAUC            \\ \hline
SPM\_SID  & 0.8663          & 0.6322          & 0.8649          & 0.6319          & 0.8400          & 0.6203          \\
DAS       & 0.8679          & 0.6352          & 0.8689          & 0.6357          & 0.8426          & 0.6237          \\
SaviorRec & 0.8687          & 0.6360          & 0.8695          & 0.6377          & 0.8435          & 0.6249          \\
COINS     & \textbf{0.8725} & \textbf{0.6394} & \textbf{0.8741} & \textbf{0.6405} & \textbf{0.8528} & \textbf{0.6301} \\ \hline
\end{tabular}
\label{tab:offline}
\end{table}

To verify the effectiveness of COINS in the click-through rate (CTR) prediction task, this section compares the performance of COINS with existing state-of-the-art methods (SPM\_SID, DAS, SaviorRec) through offline experiments, as shown in Table \ref{tab:offline}. In the full-sample (All) scenario, COINS achieves the best performance in both AUC (+0.38pp) and GAUC (+0.34pp). This proves that COINS’s overall ranking ability and resistance to user bias are significantly superior to baselines. In the warm item scenario, COINS still maintains a leading advantage, which indicates that even in scenarios with sufficient user interaction data, COINS can still optimize the representation quality of warm items and further improve prediction accuracy through the fusion of collaborative signals and semantic information in RQ-OPQ encodings. The cold-start item scenario is the core advantage area of COINS, The improvement margin is much higher than that in the full-sample and warm item scenarios. This result verifies the effectiveness of COINS’s design—by leveraging the discriminative information learning of OPQ encoding and the adaptive transfer of RQ encoding, COINS effectively solves the problem of scarce collaborative signals for cold-start items, making it the current State-of-the-Art (SOTA) method for cold-start CTR prediction tasks.

\subsection{Ablation Study}

\begin{table}[]
\caption{The ablation study resluts of COINS.}
\begin{tabular}{ccccccc}
\hline
\multirow{2}{*}{} & \multicolumn{2}{c}{All}           & \multicolumn{2}{c}{Warm}          & \multicolumn{2}{c}{Cold}          \\ \cline{2-7} 
                  & AUC             & GAUC            & AUC             & GAUC            & AUC             & GAUC            \\ \hline
only sid          & 0.8650          & 0.6312          & 0.8642          & 0.6305          & 0.8391          & 0.6192          \\
iid+sid    & 0.8671          & 0.6331          & 0.8681          & 0.6335          & 0.8401          & 0.6215          \\
iid+RQ        & 0.8702          & 0.6371          & 0.8725          & 0.6378          & 0.8479          & 0.6277          \\
iid+OPQ       & 0.8683          & 0.6345          & 0.8706          & 0.6368          & 0.8455          & 0.6257          \\
COINS   & \textbf{0.8725} & \textbf{0.6394} & \textbf{0.8741} & \textbf{0.6405} & \textbf{0.8528} & \textbf{0.6301} \\ \hline
\end{tabular}
\label{tab:ablation}
\end{table}

To verify the function of each module in COINS, we conducted sufficient ablation experiments, and the experimental results are shown in the Table \ref{tab:ablation}. Firstly, we  directly replace the original random ID in base model by RQ-OPQ encodings  (short for \textbf{only sid}). Although it achieves comparable overall performance to the base model, it exhibits inferior performance on warm items compared to the original model due to collaborative signal misalignment. Secondly, we attempted to preserve random IDs and treat SIDs as a type of ID feature (short for \textbf{iid+SID}). This variant has a slight improvement compared to the base model, as the introduction of SID brings additional information increment, but it is still far inferior to the COINS model. Finally, we ablated the two modules of COINS (short for \textbf{iid+RQ}, \textbf{iid+OPQ}). Both variants can achieve significant benefits compared to the base model. By integrating the advantages of these two variants, COINS simultaneously learns the discriminative information and convergent semantic information of items, ultimately achieving the optimal performance.

\subsection{Online A/B Testing}

\begin{table}[]
\caption{The Online A/B testing resluts of COINS.}
\begin{tabular}{ccccc}
\hline
      & \multicolumn{2}{c}{All} & \multicolumn{2}{c}{Cold} \\ \cline{2-5} 
      & Buyer    & Order Volume & Buyer     & Order Volume \\ \hline
base  & -        & -            & -         & -            \\
COINS & +1.720\% & +2.230\%     & +3.512\%  & +9.639\%     \\ \hline
\end{tabular}
\label{tab:online}
\end{table}
To verify the efficiency of COINS's online platform, we compared it with an online ranking model in the e-commerce mall search scenario through rigorous A/B testing. We conducted a 14 day experiment under real online traffic, with 5 days being the AA period and the remaining 9 days being the AB period. The base group and exp group were randomly assigned 7\% of the overall real online traffic. The experimental results are shown in the Table \ref{tab:online}. Our experimental focus indicators are the \textbf{buyers} and \textbf{order volume}, and we also report the relevant indicators in the cold start scenario in the Table \ref{tab:online}. Experiments have shown that COINS can increase the overall number of buyers by 1.72\% and the order volume by 2.23\%. Simultaneously, COINS significantly improves the cold start scenario (buyer number +3.512\% and order volume +9.639\%) in mall ecommerce search.

\section{Conclusion}
To address cold-start challenges in e-commerce, we propose COINS: it fuses item ID collaboration and semantics RQ via adaptive transfer, and enhances item differentiation with OPQ contrastive learning. Offline experiments and online A/B testing confirm its effectiveness (improving buyer +1.72\%, order volume +2.23\%). In future work, we will explore the fusion of multimodal semantics and collaborative signals to adapt to more complex cold-start scenarios.

\bibliographystyle{ACM-Reference-Format}
\bibliography{sample-base}

% \appendix









\end{document}